# Oxygen vacancy enhanced room temperature ferromagnetism in Al-doped MgO nanoparticles

Debabrata Mishra, Balaji P. Mandal, Rupam Mukherjee, Ratna Naik, Gavin Lawes, and Boris Nadgorny

*Department of Physics and Astronomy, Wayne State University, Detroit MI, 48201 USA*

**Abstract**

We have investigated the room temperature ferromagnetic order that develops in Al-substituted magnesium oxide, Mg(Al)O, nanoparticles with Al fractions of up to 5 at.%. All samples, including undoped MgO nanoparticles, exhibit room temperature ferromagnetism, with the saturation magnetization reaching a maximum of 0.023 emu/g at 2 at.% of Al. X-ray photoelectron spectroscopy identifies the presence of oxygen vacancies in both doped and undoped MgO nanoparticles, with the vacancy concentration increasing upon vacuum annealing of Mg(Al)O, resulting in two-fold enhancement of the saturation magnetization for 2 at.% Al-doped MgO. Our results suggest that the oxygen vacancies are largely responsible for room temperature ferromagnetism in MgO.

Over the past decade, there has been tremendous interest in understanding the origin of room temperature unconventional ferromagnetism found in a variety of semiconducting and insulating oxides.[1,2] The observation of a weak ferromagnetism in insulating oxides was particularly puzzling. This ephemeral magnetism has been attributed to a variety of mechanisms, including intrinsic defects, secondary phases and inadvertent contamination[3]. While Sundaresan *et al.* have argued that ferromagnetism is a near-universal, intrinsic feature of wide band gap oxide nanoparticles[2], extrinsic contamination can be exceedingly difficult to exclude conclusively.[4] Nevertheless, regardless of the microscopic origin of this magnetism, routine observations of a small ferromagnetic signal in nominally nonmagnetic oxides are indisputable.

Most of the studies in this field have been focused on oxide and nitride dilute magnetic semiconductors including $In_2O_3$, $TiO_2$, ZnO, and InN thin films because of their potential application in spintronics devices.[5] A number of different mechanisms may contribute to the development of the ferromagnetism, including effects arising from paramagnetic transition metal ions, impurities, point defects and dislocations, changes in carrier concentration, and grain boundaries[3]. Theoretical computations based on *ab initio* methods suggest the possibility of carrier mediated ferromagnetism in these dilute magnetic semiconductors.[6,7] The experimental verification of carrier controlled room temperature ferromagnetism in transition metal doped oxides, such as in Cr-doped $In_2O_3$, supports this proposal.[8] More recent experiments suggest a connection between room temperature ferromagnetic order and a spontaneous spin polarization of the conduction electrons, even in the absence of magnetic dopants[9]. While the origin of room temperature ferromagnetism in wide band gap oxides/insulators such as CaO, $Al_2O_3$, MgO is still hotly debated, various defects, specifically oxygen vacancy defects and mobile charge carriers are believed to have an effect on the ferromagnetic order in semiconductors.[10]

Magnetism in transparent MgO thin films and nanoparticles has been investigated by a number of different groups.[11-15] This room temperature ferromagnetism in MgO is particular interesting because large band gap for MgO (~7.8 eV compared to 2-3 eV for semiconducting oxides) drastically modifies any possibly electronic contribution to the magnetism. MgO is a technologically important material, used in magnetic tunnel junctions, and can also serve as an attractive model system to study the defect-induced magnetic properties because of its simple rock salt structure.[16,17] For example, recent theoretical studied have considered the substitution of nitrogen (or carbon) for oxygen in MgO, which may form nitrogen pair defects, resulting in ferromagnetic interaction in MgO.[18,19] In MgO nanoparticles prepared by chemical synthesis, the ferromagnetism has been attributed to Mg vacancies at the grain boundaries.[12] In contrast to this cation vacancy model, Maoz *et al.*[20] have suggested that unpaired electrons trapped at the oxygen vacancy sites may be responsible for the magnetism in a highly defective MgO nanosheet.

Nanoparticles, with their large surface-to-volume ratio can be especially effective in producing various structural defects. In this Letter, we report on the correlation between composition, local structure, and magnetic properties of Al-doped MgO nanoparticles. We show that X-ray photoemission spectroscopy (XPS) is sensitive to the formation of oxygen vacancies, produced by Al doping and vacuum annealing process. The changes in the XPS oxygen spectra are likely to be associated with different oxygen coordination numbers, allowing us to establish a correlation between oxygen concentration and magnetization in MgO nanoparticles. While the interaction with Mg point defect vacancies and Al dopants may also contribute to the development of room temperature ferromagnetism, our results provide direct evidence for oxygen vacancy-enhanced magnetic properties in Al-doped MgO.

In order to prepare the Al-substituted MgO nanoparticles, we started with $Mg(NO_3)_2 \cdot 6H_2O$ and $Al(NO_3)_3 \cdot 9H_2O$ nitrates. They were mixed in stoichiometric ratio in an aqueous solution, which was then heated at $400^0$ C to form nanocrystalline powders then subsequently calcined at 550 °C for 45 min. Some samples were vacuum annealed by heating to 600 °C for six hours in a background pressure of $10^{-6}$ Torr. The phase purity of the samples was verified by x-ray powder diffraction (XRD), using a Rigaku RU2000 rotating anode powder diffractometer. In addition, the microstructure of both the doped and undoped MgO nanoparticles was characterized using high resolution electron microscopy. A chemical analysis of the as-prepared and annealed samples was performed using X-ray photoelectron spectroscopy (XPS). The magnetic properties were characterized using a Quantum Design SQUID magnetometer. We confirmed the sample expected doping levels using SEM EDX.

Figure 1(a) shows the x-ray diffraction pattern of the as-prepared, 2% and 5 % Al substituted MgO nanoparticles. The characteristic MgO peaks are labeled. No additional peaks are observed, which indicates that no crystalline impurity phases are present. All nanoparticles were nearly spherical; the average crystalline size calculated using the Debye-Scherrer formula was found to be approximately 10 nm for all samples. There is a shift in peak positions associated with Al substitution, which implies that the Al dopants are incorporated into the MgO lattice to form a solid solution. To parameterize the particle size and grain boundary region of the nanoparticles, we recorded transmission electron microscope (TEM) images for all samples. Figure 1(b) shows the dark field image of Al:MgO for 2 at. % Al substitution. Our results confirm that the particles are spherical with no evidence for secondary phases. To probe the grain boundary region more carefully, we took high resolution TEM images, shown in Fig. 1(c). The arrows indicate the lattice planes of MgO. The inset in Fig. 1(c) highlights the lattice distortion in the MgO planes produced by Al substitution. These pictures strongly suggest that the point defects in Al-substituted MgO produce structural changes near grain boundaries.

We used energy dispersive x-ray spectroscopy (not shown) to analyze the elemental composition of the samples and to look for the presence of magnetic transition metal contaminants, which could affect the magnetic properties. Within the resolution of our spectrometer, we did not observe any evidence for contamination in the samples. All possible precautions were used to avoid accidental contamination: only non-metallic tweezers were used, and a number of control tests were performed, including measurements of the Raman spectra, as these are sensitive to amorphous magnetic impurity phases, such as magnetite.[21] We did not find any traces of magnetic impurity phases from these more detailed studies.

As a further probe of the chemical structure of these samples, we measured X-ray photoemission spectra (XPS) for doped and undoped MgO samples prior to and subsequent to vacuum annealing. A representative curve for the Al (2 at.%) doped MgO nanoparticles is shown in Fig. 2(a). All the peaks can be indexed to Mg, Al, and O, together with residual C on the surface, and we did not find any additional peaks after vacuum annealing. This argues against accidental contamination of the samples as a result of vacuum annealing process. To quantitatively compare the oxygen vacancy concentration in the as-prepared and vacuum annealed doped and undoped MgO samples, we fitted the O 1s peaks, as shown in Fig. 2(b). We find asymmetric Gaussian peaks in both doped and undoped MgO samples, with the degree of asymmetry increasing after vacuum annealing. The largest asymmetry is observed in the 2 at.% Al doped MgO nanoparticles, both before and after vacuum annealing. The asymmetry in the O 1s peaks suggests that multiple O valences are present in these samples.

In order to quantitatively explore the different O coordination in these samples, we fitted the peaks using three separate Gaussian curves having binding energies at 530.2 eV, 531.8 eV and 533.3 eV, corresponding to the O binding energy of MgO, oxygen vacancies, and binding in surface $O_2$, respectively.[22, 23] The results, determined by fitting the curves, are plotted in Fig. 2(c). These show that oxygen vacancies are present in the as prepared particles for both doped and undoped MgO samples. In the as-prepared undoped MgO nanoparticles, we find that the signal attributed to oxygen vacancies contributes 4.8% to the total, increasing to 6% for 2 at.% Al:MgO but then decreasing with further Al substitution. Although the XPS signal depends on the scattering efficiencies, so that these estimates may not precisely reflect the actual oxygen vacancy fraction in the samples, these results provide insight into the relative change in the concentration of these defects among the different samples included in this study. The variation in the oxygen vacancy signal with Al fraction, for nanoparticles having the same size, suggests that oxygen vacancies may be synergistically affected by other point defects in these nanoparticles. After vacuum annealing the amplitude of the oxygen vacancy signal nearly doubles for both doped and undoped MgO samples, reaching a maximum of 12 % of the fitted peak, for 2at.% Al doped MgO. Our results are similar with those reported by Yang *et al*, who found 10% oxygen vacancy in $Al_2O_3$ nanoparticles.[23]

Figure 3 (a) plots the room temperature magnetization *M* versus magnetic field *H* for the undoped and Al-doped MgO nanoparticles, after correcting for the diamagnetic contribution estimated from the high-field magnetization. The saturation magnetization $M_S$ values for each composition have been plotted in Fig. 3 (b). Examining Fig. 3, we observe that $M_S$ increases from 0.009 emu/g for pure MgO, to 0.023 emu/g at 2 at% of Al substitution, with a systematic decrease in magnetization with a further increase in Al fraction. The systematic variation of the $M_S$ with Al content provides confirmation that the magnetic signal does not arise from accidental impurity phases, which would not be expected to be correlated with the Al composition.

To verify the critical role of oxygen vacancies, we investigated the magnetic properties of these nanoparticles upon air and vacuum annealing to vary the oxygen vacancy concentrations for a given Al composition. To test the stability of the defect-induced ferromagnetic order in the as-prepared (AP) nanoparticles, we measured the magnetic properties before and after annealing at high temperature in air. The magnetization loops for nanoparticle sample annealed in air at 1000 °C for 12 hours show only a minimal change in magnetic properties compared to the as-prepared samples. Representative curves for the 2 at. % Al- substituted sample are shown in Fig. 4(a). This suggests that the defects producing the ferromagnetism are robust, which has important implications for the potential device applications for these materials.

Conversely, we find that vacuum annealing (at 600 °C for 6 hours at $10^{-6}$ torr) these Al-substituted MgO nanoparticles samples, to introduce additional vacancies, produces a remarkable change in the magnetic properties, as shown in Fig. 4. Fig. 4(a) plots the room temperature magnetization loops measured for as-prepared (AP), air annealed (AA), and vacuum annealed (VA) $Mg_{1-x}Al_xO$ for $x = 0.02$ and Fig. 4(b) plots the saturation magnetization versus Al fraction for the AP, AA and VA nanoparticles. The $M_S$ value increases from 0.009 and 0.025 emu/g to 0.013 emu/g and 0.046 emu/g for the pure MgO and 2 at.% Al-doped MgO nanoparticles respectively. These results suggest that oxygen vacancy defects lead to the increase of the magnetization and, furthermore, that the relative magnitude of this increase depends on the Al content. This is similar to the results seen in insulators like $HfO_2$ powders[24], and strongly suggests some synergistic interaction among the Al dopants and oxygen vacancy defects. Such interaction effects are also supported by our XPS results, which find a maximum oxygen vacancy concentration for the 2 at.% Al doped MgO nanoparticles (see Fig. 2(c)). The small change in the magnetization observed for undoped MgO is consistent with the earlier result reported by Hu *et al*,[11] and suggests that the ferromagnetism in pure MgO is not particularly sensitive to the concentration of oxygen vacancy defects. Conversely, we observe a much larger relative increase in the magnetization of the vacuum annealed Al (2 at.%) doped MgO.

While the importance of oxygen vacancies in the development of ferromagnetism in MgO nanoparticles is well-established, and confirmed by these studies, the effects of Al doping on magnetic order are less transparent. One suggestion is that the Al dopants produce lattice distortions and structural changes near grain boundaries (as seen in Fig. 1c), which could reduce the energy of oxygen vacancy formation at low concentrations of Al. This is consistent with the fact that the maximum of oxygen concentration at ~ 2.5% of Al is more pronounced for vacuum annealed than for as prepared nanoparticles. However, these Al ions could also affect the ferromagnetic structure in the MgO nanoparticles through other mechanisms. At low concentrations, Al ions, which enter the MgO lattice as substitutional defects in the 3+ valence, could increase the number of Mg vacancies, which as some studies have suggested[20] may enhance the development of ferromagnetism in MgO by trapping free electrons at the oxygen vacancy site. This could explain the increase in magnetization in the Al substituted nanoparticles. It has been suggested by *ab initio* band structure calculations on the wide band gap insulator CaO that a small concentration of Ca vacancies, less than 3%, can induce ferromagnetism.[25] In this model, the point defects will initially bind the charge carriers at neighboring sites and produce local magnetic moments. On the other hand, unlike adjacent magnetic dopants, which usually promote antiferromagnetic coupling,[26] the effects of clustering in the development of $d^0$ ferromagnetism can be more subtle.[19,20] Indeed, the probability of having an isolated Al defect in MgO is almost independent of the Al fraction within our concentration range (between $0 < x < 0.05$), as it decreases from 100% to about 75% at $x = 0.05$, while the probability of having two adjacent Al defects (dimers) increases from 5.5% at $x = 0.01$ to nearly 20% at x = 0.05, with a similar increase in the probability of having triplet Al defects in this range of concentrations[27]. The non-monotonic dependence of the saturation magnetization on Al fraction may be ascribed to a combination of increasing oxygen vacancy defects stabilized by Al dopants combined with the competing effects of monomer, dimer, and trimer Al defects suppressing ferromagnetism.

We summarize the results of our study in Figure 4(b). For all treatment conditions, including the as-prepared, high temperature air annealed, and vacuum annealed samples, the magnetization exhibits a maximum for the 2.5 at.% Al-substituted MgO nanoparticles. The fact that the maximum saturation magnetization occurs at 2.5 at.% Al for all samples implies that the mechanism through which Al doping contributes to the ferromagnetic order does not depend on the oxygen vacancy defect concentration.[28] At the same time, the fact that the maximum of the saturation magnetization coincides with the maximum concentration of oxygen vacancies for both as prepared and oxygen-annealed samples, strongly indicate that oxygen vacancies play a key role in the development of the ferromagnetic order.

In summary, we have investigated the effect of Al substitution (between 0 and 5 at.%) on the room temperature ferromagnetic order of MgO nanoparticles with different oxygen vacancy concentration. The oxygen vacancy fraction increases with Al doping in the as-prepared particles, with a further increase on vacuum annealing, reaching a maximum for the 2 at.% Al substituted MgO. The largest $M_S$ value were observed for ~2.5 at.% Al samples over all post-preparation treatments, with the moment decreasing with higher Al fractions. Vacuum annealing produces only a modest increase in the saturation magnetization for the undoped MgO nanoparticles, but dramatically enhances $M_S$ value for the Al substituted MgO, with the largest relative increase occurring for 2.5 at.% Al. Using the XPS spectra analysis, we have established the direct correlation between the oxygen vacancy concentrations and the values of magnetization of MgO nanoparticles. While we emphasize the importance of the interplay of different defects, our results highlight the dominant role of oxygen vacancies in the development of room temperature ferromagnetism in MgO. This work was supported by the National Science Foundation under DMR-1006381.

**References:**

[1]M. Venkatesan, C. B. Fitzgerald, and J. M. D. Coey, Nature **430**, 630 (2004).

[2]A. Sundaresan, R. Bhargavi, N. Rangarajan, U. Siddesh, and C. N. R. Rao, Phys. Rev. B **74**, 161306 (2006).

[3]J. Coey, Solid State Sci. 7, 660 (2005); O. Volnianska and P. Bogusławski, J. Phys. Condens. Matter 22, 073202 (2010).

[4]D. W. Abraham, M. M. Frank, and S. Guha, Appl. Phys. Lett. **87**, 252502 (2005).

[5]N. H. Hong, J. Sakai, N. Poirot, and V. Brizé, Phys. Rev. B **73**, 132404 (2006); Q. Wang, Q.Sun, G. Chen, Y. Kawazoe, and P. Jena, Phys. Rev. B **77**, 205411 (2008).

[6]A. Walsh, J. L. F. Da Silva, and S.-H. Wei, Phys. Rev. Lett. **100**, 256401 (2008).

[7]K. Sato and H. Katayama-Yoshida, Semiconductor Science and Technology **17**, 367 (2002).

[8]J. Philip, et al., Nat Mater **5**, 298 (2006).

[9]R. P. Panguluri, P. Kharel, C. Sudakar, R. Naik, R. Suryanarayanan, V. M. Naik, A. G. Petukhov, B. Nadgorny, and G. Lawes, Phys. Rev. B **79**, 165208 (2009).

[10]H. S. Hsu, et al., Appl. Phys. Lett. **88**, 242507 (2006).

[11]J. Narayan, S. Nori, D. K. Pandya, D. K. Avasthi, and A. I. Smirnov, Appl. Phys. Lett. **93**, 082507 (2008).

[12]J. Hu, Z. Zhang, M. Zhao, H. Qin, and M. Jiang, Appl. Phys. Lett. **93**, 192503 (2008).

[13]C. M. Araujo, et al., Appl. Phys. Lett. **96**, 232505 (2010).

[14]H.-X. Gao and J.-B. Xia, J. Appl. Phys. **111**, 093902 (2012).

[15]A. M. H. R. Hakimi, M. G. Blamire, S. M. Heald, M. S. Alshammari, M. S. Alqahtani, D. S. Score, H. J. Blythe, A. M. Fox, and G. A. Gehring, Phys. Rev. B **84**, 085201 (2011).

[16]J. Mathon and A. Umerski, Phys. Rev. B **63**, 220403 (2001).

[17]C. Århammar, C. Moyses Araujo, K. V. Rao, S. Norgren, B. Johansson, and R. Ahuja, Phys.Rev. B **82**, 134406 (2010).

[18]H. Wu, A. Stroppa, S. Sakong, S. Picozzi, M. Scheffler, and P. Kratzer, Phys. Rev. Lett. **105**, 267203 (2010).

[19]B. Gu, N. Bulut, T. Ziman, and S. Maekawa, Phys. Rev. B **79**, 024407 (2009).

[20] B.M. Maoz, E. Tirosh, M. Bar Sadan, and G. Markovich, Phys. Rev. B **83**, 161201 (2011).

[21]L. Gasparov, A. Rush, T. Pekarek, N. Patel, and H. Berger, J. Appl. Phys. **105**, 07E109 (2009).

[22]J. S. Corneille, J.-W. He, and D. W. Goodman, Surf. Sci. **306**, 269 (1994).

[23]G. Yang, D. Gao, J. Zhang, J. Zhang, Z. Shi, and D. Xue, J. Phys. Chem. C **115**, 16814 (2011).

[24]J. M. D. Coey, M. Venkatesan, P. Stamenov, C. B. Fitzgerald, and L. S. Dorneles, Phys. Rev. B **72**, 024450 (2005).

[25]I. S. Elfimov, S. Yunoki, and G. A. Sawatzky, Phys. Rev. Lett. **89**, 216403 (2002).

[26]G. Lawes, A. S. Risbud, A. P. Ramirez, and R. Seshadri, Phys. Rev. B **71**, 045201 (2005).

[27]R. E. Behringer, J. Chem. Phys. **29**, 537 (1958).

[27]J. M. D. Coey, M. Venkatesan, P. Stamenov, C. B. Fitzgerald, and L. S. Dorneles, Phys. Rev. B **72**, 024450 (2005).

[28]X. G. Xu, et al., Appl. Phys. Lett. **97**, 232502 (2010).

**Figure Captions:**

FIG.1 (a) X-ray diffraction patterns for $Mg_{1-x}Al_xO$ nanoparticles. (b) Dark field TEM image and (c) HRTEM image of 2 at.% Al doped MgO nanoparticle.

FIG. 2 (Color Online) (a) Wide scan x-ray photo-electron spectra for as-prepared and vacuum annealed (at $600^0$ C for 6 hours) 2 at.% Al doped MgO nanoparticles (b) x-ray photo-electron spectra of O 1S peak of (i-iii) as-prepared and (iv-vi) vacuum annealed pure , 2 and 5 at.% Al doped MgO respectively (Colored lines inside the main Gaussian peak are fittings to different oxygen coordination.) and (c) Oxygen vacancy concentration of as-prepared and vacuum annealed pure , 2 and 5 at.% Al doped MgO.

FIG. 3 (Color Online) a) $M$ - $H$ loop and (b) Saturation magnetization $M_S$ as as a function of $x$ in $Mg_{1-x}Al_xO$ samples at 300K. (The dotted line is a guide to the eye.)

FIG. 4 (Color Online) (a) $M$ - $H$ loop (at 300 K) for as prepared (AP), air annealed (AA), and vacuum annealed (VA) $Mg_{1-x}Al_xO$ samples (for $x$=0.02); (b) $M_S$ as a function of $x$ for AP, AA, and VA $Mg_{1-x}Al_xO$ samples. (The solid line shows a spline interpolation of the curve between the measured data points.)

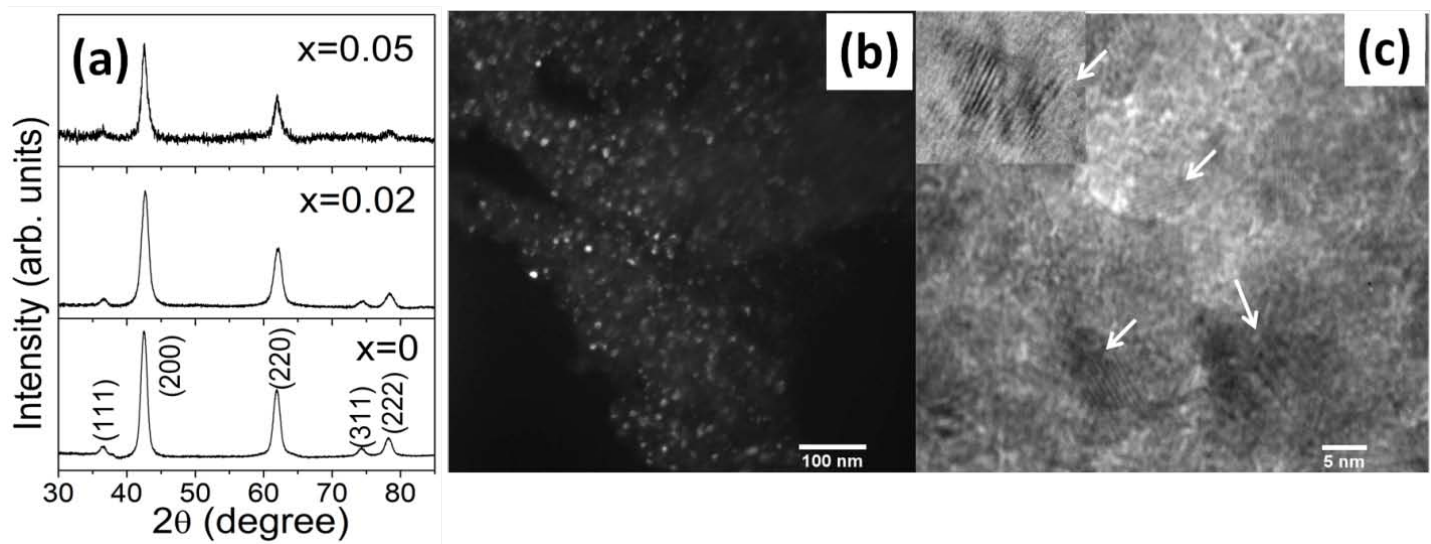

Figure 1

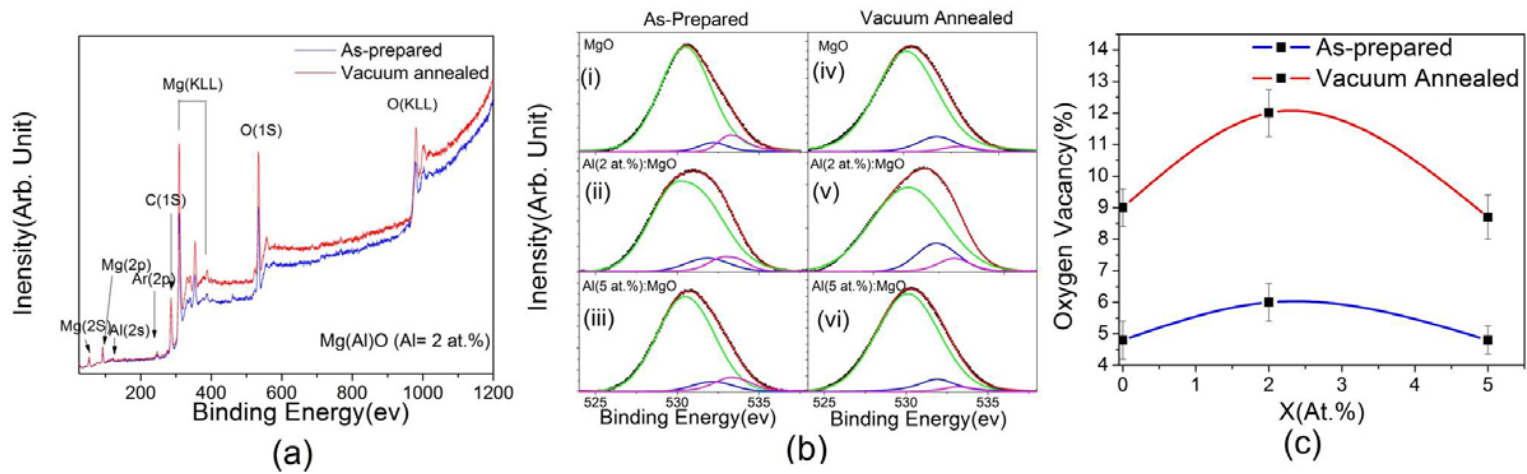

Figure 2

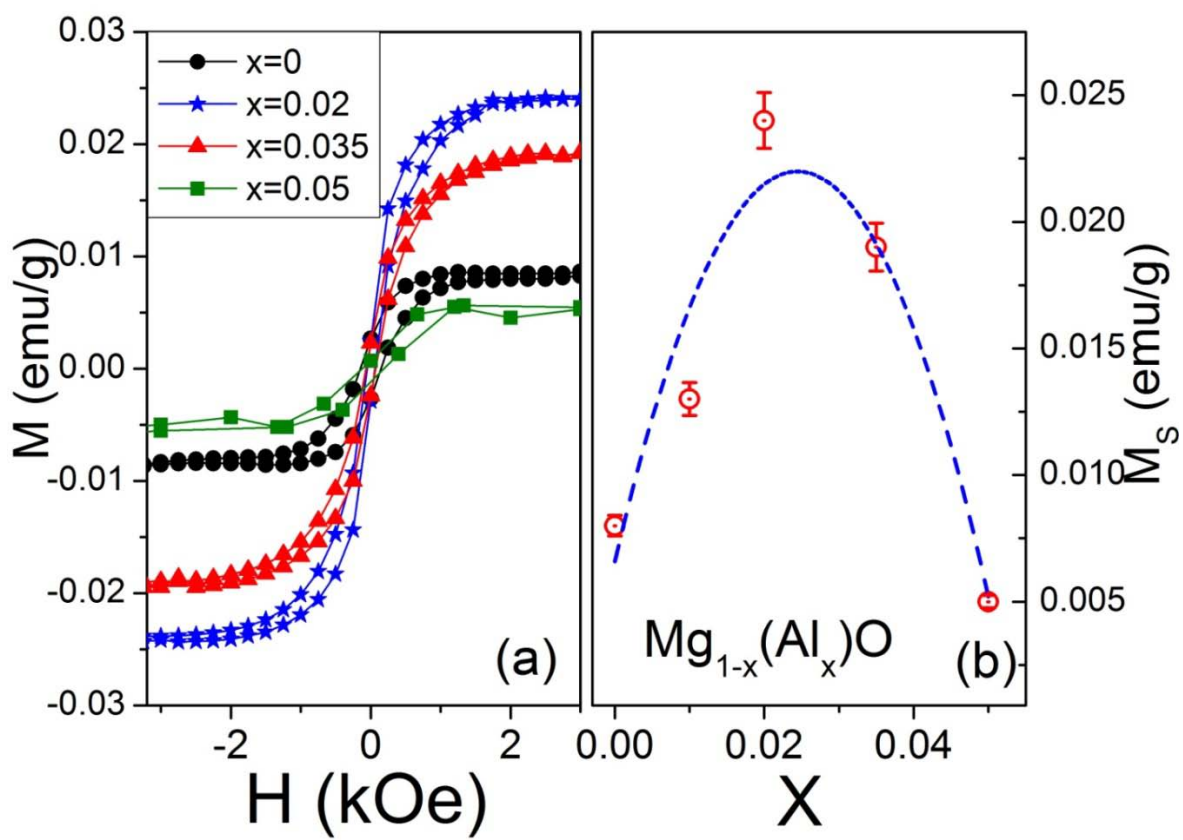

Figure 3

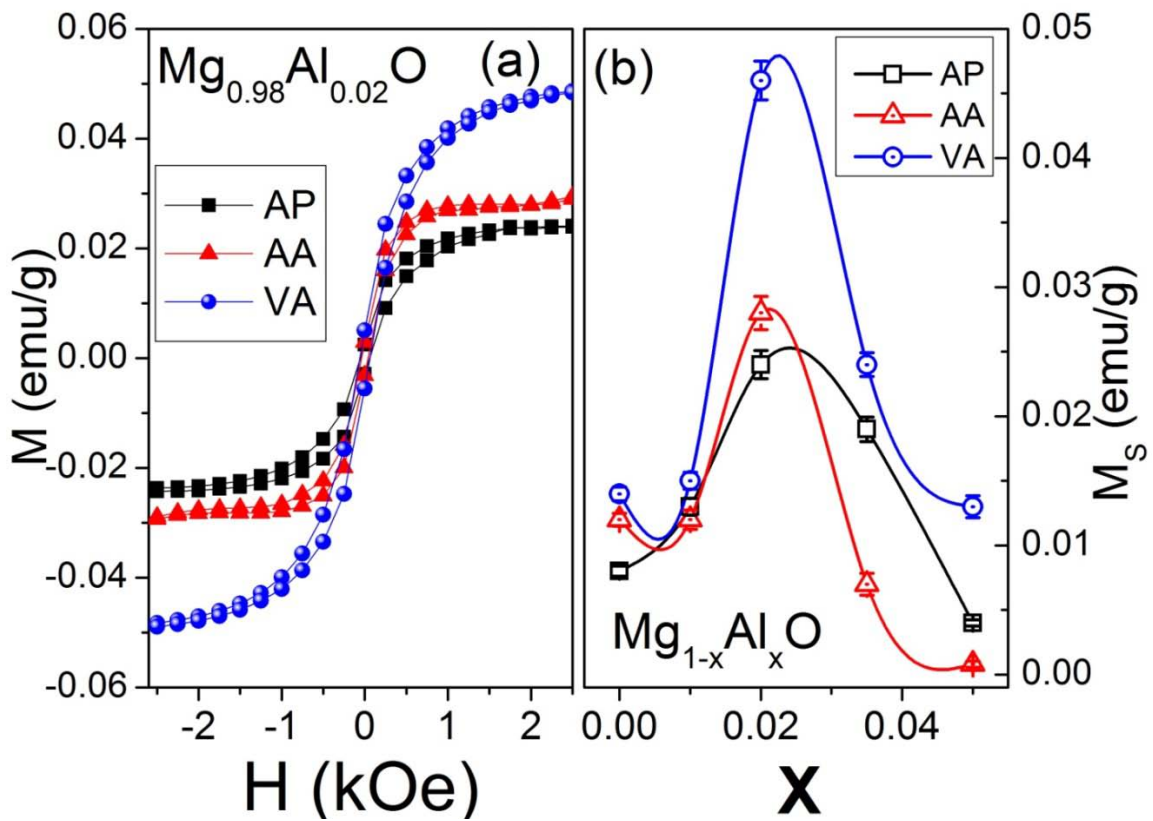

Figure 4